\documentclass[12pt]{article}
\newcommand{\beq}{\begin{equation}}
\newcommand{\eeq}{\end{equation}}
\newcommand{\bea}{\begin{eqnarray}}
\newcommand{\eea}{\end{eqnarray}}
\newcommand{\epo}{e^+_1}
\newcommand{\emo}{e^-_1}

\newcommand{\Emo}{E^-_1}

\newcommand{\omo}{\omega_1}
\newcommand{\pz}{\partial_0}
\newcommand{\po}{\partial_1}
\newcommand{\pzinv}{\frac{1}{\partial_0}}
\newcommand{\pzinvsq}{\frac{1}{\partial^2_0}}

\newcommand{\prop}{\Theta}
\def\subdef#1{\gdef\globalColor##1{##1}}      
      \subdef{Black}

\newcommand{\plabel}{\label}
\usepackage{epsfig}
\usepackage{amstex}
\usepackage{cite}

\begin{document}

\begin{titlepage}
\renewcommand{\thefootnote}{\fnsymbol{footnote}}

\epsfig{file=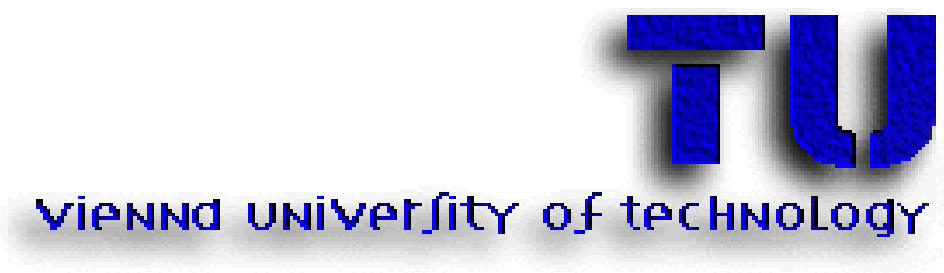,height=0.8cm}
\hfill TUW-97-08 \\
\begin{center}
\vspace{1cm}

\textbf{\Large Nonperturbative path integral of 2d dilaton gravity 
and two-loop effects from scalar matter}

\vfill
\renewcommand{\baselinestretch}{1}

{\bf W. Kummer$^1$\footnotemark[1], H. Liebl$^1$\footnotemark[2]
       and D.V. Vassilevich$^{1,2}$\footnotemark[3]}

\vspace{7ex}

{$^1$Institut f\"ur
    Theoretische Physik \\ Technische Universit\"at Wien \\ Wiedner
    Hauptstr.  8--10, A-1040 Wien \\ Austria}

\vspace{2ex}

{$^2$Department of Theoretical Physics \\
   St. Petersburg University \\ 198904 St. Petersburg \\  Russia}

  \footnotetext[1]{e-mail: \texttt{wkummer@@tph.tuwien.ac.at}}
  \footnotetext[2]{e-mail: \texttt{liebl@@tph16.tuwien.ac.at}}
  \footnotetext[3]{e-mail: \texttt{vasilev@@snoopy.niif.spb.su}}

\end{center}
\vfill

\begin{abstract}
Performing an nonperturbative path integral for the geometric
part of a large class of 2d theories without kinetic term for
the dilaton field, the quantum effects from scalar matter fields
are treated as a perturbation.  When integrated out to two-loops
they yield a correction to the Polyakov term which is still exact
in the geometric part.  Interestingly enough the effective
action only experiences a renormalization of the dilaton
potential. 
\end{abstract}
\end{titlepage}
\vfill

\section{Introduction}

Gravity in four dimensions still resists successful quantization. 
Also many quantum properties of the black 
hole and its interactions with quantized matter \cite{haw75} have not 
found a generally accepted final clarification. 
In recent
years however, remarkable progress has been made in the simplified 
setting of two dimensional models, including models with a 
dilaton field beside the metric \cite{lou94a}, but also models with torsion 
\cite{kat86}. In fact, by local transformations of the fields a theory 
with torsion may be transformed into a dilaton theory 
\cite{kat96}. 
Then also the dilaton field may be finally transformed away 
\cite{lou94a}. However, it is not correct that as a consequence of this 
line of argument a quantization of {\it any} theory in d = 2 may 
be achieved by quantizing theories depending on the metric alone 
by elimination of the dilaton field \cite{kuc97}: Profound changes of 
the (classical) global structure tied to the transition of such 
theories with respect to a conformally related dilaton theory 
\cite{kat97} indicate that equally drastic changes must be expected 
in the respective quantum theories. \\

At present three basic approaches for the quantum theory of 2d 
gravity models are known. The first one uses the particular 
structure of the theory of constraints to find the explicit 
canonical solution yielding --- in the absence of matter-fields 
--- a finite dimensional phase space 
\cite{kuc97},\cite{lou94b,can94,ben94,thi93,kuc94,str94,sch94}. However, an 
extension of that result so as to include interactions of the 
geometric variables of matter seems difficult. A second one 
emerged as a by-product of string theory. The corresponding
dilaton theory containing black hole solutions 
can be solved classically in the presence of 
matter \cite{man91}, but the hope that conformal field theory 
techniques suffice for a complete quantization could not be 
fulfilled. Thus also for that model only 
semiclassical studies were possible. 

The prejudice that path-integral quantization is necessarily 
related to perturbation theory seems to have prevented its 
nonperturbative application to quantize 2d covariant theories 
until recently. This third approach is based upon an earlier  
observation that the path integral for the model of \cite{kat86} could 
be performed exactly in a suitable gauge \cite{kum92}. However, the 
present authors also realised that this remains even true for all 
matterless theories, whether dilaton fields are present or not 
\cite{kum97}. It would have been difficult to find this simple result 
from the involved perturbative calculations in the background 
field gauge \cite{odi92}. Thus full agreement with the first approach 
was confirmed in the sense that no local quantum effects appear.

Quantum effects in 1+1 gravity without matter
are restricted to global fluctuations, e.g. zero modes of compactified
directions \cite{sch94}. These we exclude by assuming that the 
path integral is evaluated on spacetimes
with simply connected topology.
However, when additional matter fields are added also the method of 
\cite{kum97} failed, except for the particular
case of the Jackiw-Teitelboim model \cite{tei83}.
What can be done, though, is to consider  perturbation theory in the 
{\it matter field},
treating the geometrical part still exactly by a {\it nonperturbative} 
path integral. Our approach thus differs fundamentally from the conventional
`semiclassical' one \cite{man91} in which (mostly only one loop) effects of
matter are added and the resulting effective action subsequently is solved
classically. It also clearly aims to go beyond the consideration 
of the backreaction between in- and outgoing matter \cite{kie95} in a 
fixed background.

In the present work we compute the generating functional for connected Green
functions as well as the effective action for a wide
class of 2d gravity models minimally coupled to scalar matter
to two loop order (Section \ref{sec-two-loop}). 
The gravitational part of the theory is treated exactly in
the sense that no loops consisting only of gravitational 
propagators appear at all. 
We demonstrate the remarkable fact that the two loop contribution of 
the effective action vanishes.
More precisely, the only effect of two--loop corrections is a renormalization
of the dilaton potential.
It follows that any one loop result of 2d gravity contained
in our class of models is automatically true to two loops.
As in the previous paper \cite{kum97}, our analysis is local.
This means that we suppose  asymptotic fall-off conditions for
quantum fields and neglect all surface terms which may appear 
when integrating  by parts.

Recently an attempt to analyse two--loop corrections in dilatonic
models has been made by Mikovic and Radovanovic \cite{mik96} 
who considered the CGHS model minimally coupled to matter.
Strictly speaking, their model does not fall into the class
considered in the present paper. Some implications of the
calculations of that reference are discussed in the conclusion 
(Section \ref{sec-concl}).

\section{Two Loop Quantization}
\label{sec-two-loop}

Our starting point is  the action for 1+1
gravity with a spacetime integral over the Lagrangian 
\begin{equation}
\plabel{sec2a}
{\cal{L}}={\cal{L}}^g + {\cal{L}}^m +{\cal{L}}^s \quad ,
\end{equation}
which is a sum of the gravitational, the matter and a source contribution.
We will work with a gravitational term in its 
 first order form \cite{gros92}
\begin{equation}
  \plabel{lfirst}
  {\cal{L}}_{(1)}=X^+De^-+ X^-De^++Xd\omega +\epsilon (X^+X^-U(X) 
  + V(X))\; ,
\end{equation}
where $De^a=de^a+(\omega \wedge e)^a$ is the torsion two form,
the scalar curvature
$R$ is related to the spin connection $\omega$ by $-\frac{R}{2}=* d\omega$
and $\epsilon$ denotes the volume two form
$\epsilon=\frac{1}{2}\varepsilon_{ab}e^a \wedge e^b=d^2x   \det
e_a^{\mu}=d^2x\,(e)$.  Our conventions are determined by $\eta=diag(1,-1)$
and $\varepsilon ^{ab}$ by $\varepsilon ^{01}=-\varepsilon ^{10}=1$. We
also have to stress that even with Greek indices $\varepsilon^{\mu \nu}$
 is always understood to be the antisymmetric Levi-Civita symbol and never 
the corresponding tensor. 
In \cite{kum97} it was shown that (\ref{lfirst}) 
even at the quantum level is equivalent to the more
familiar second order form
\begin{equation}
 \plabel{lbegin}
{\cal{L}}_{(2)}=\sqrt{-g}
\left(-X\frac{R}{2}-V(X) + U(X)\, (\nabla X)^2\, \right)\; .
\end{equation}

In many studies  an exponential parametrization for the dilaton field
$X=e^{-2\phi}$ is chosen which restricts $X$ to ${\Bbb{R}_+}$ and from the
global point of view represents a further drastic assumption \cite{kat96}.
Our matter contribution is a minimally coupled scalar field whose 
Lagrangian ${\cal{L}}^m$ is given by 

\begin{equation}
\plabel{lmatter}
{\cal{L}}^m = \frac12 \sqrt{-g}g^{\mu \nu}\partial_{\mu}S \partial_{\nu}S 
=- \frac12 \frac{\varepsilon^{\alpha \mu} \varepsilon^{\beta \nu}}{e}
\eta_{ab}e^a_{\mu}e^b_{\nu}\partial_{\alpha}S \partial_{\beta} S \quad .
\end{equation}
The most general dilaton gravity action (\ref{lbegin}) contains the term
$U(X)(\nabla X)^2$. This term can be removed by a dilaton
dependent conformal redefinition of the metric. The matter
action (\ref{lmatter}) is invariant under such a redefinition. However,
the quantum theory is changed: The source terms in eq. (\ref{source}) 
below acquire field dependent (conformal) factors, destroying straightforward
quantum integrability. In addition the
path integral measure for the scalar field is
changed. Thus, in this section, in a first step we restrict ourselves to a
subclass of 2d models with $U(X) = 0$ to avoid further 
technical complexities. 
Of course, in this way realistic models like spherically symmetric 4d general
relativity \cite{tho84}  $U(X) \propto X^{-1}$ are eliminated.

The same class of models with $U(X) = 0$ in (\ref{lbegin}) has also been studied in
\cite{lou94a,lou94b}.
These theories still contain a vast variety of models allowing de facto
an arbitrary topological structure which may be designed at will following a 
simple set of rules \cite{klo96a,klo96b}. Recently a particular example of
a black hole solution determined by $V(X) \propto \exp (X)$ has been advocated 
by Cruz et al.\ \cite{cru96} which reflects the same classical features as the 
CGHS model \cite{man91}.

The Lagrangian ${\cal{L}}_s$ containing the source terms for our fields is
given by
\begin{equation}
  \plabel{source}
  {\cal{L}}_s = j^+e_1^- + j^-e_1^+ + j\omega_1 +J^+X^- +J^-X^+ +JX 
  +QS\; .
\end{equation}
Using an Eddington-Finkelstein gauge for the metric defined by
a temporal (Weyl type) gauge for the Cartan variables
\begin{equation}
  \plabel{gaugefix}
  e_0^+ = \omega_0=0 \quad , \quad e_0^-=1
\end{equation}
yields the trivial Faddeev-Popov determinant \cite{kum97} 
\begin{equation}
F=(\det \partial_0 )^3 \quad . 
\plabel{FPdet}
\end{equation}
In this gauge the actions (\ref{lfirst}) and (\ref{lmatter}) are
\bea
\plabel{lfirst-gf}
{\cal L}^g_{gf}&=& X^+\pz \emo + X^- \pz \epo +X\pz \omo +X^+ \omo -\epo V(X) \\
{\cal L}^m_{gf}&=& \left( \emo (\pz S)^2 -(\pz S)(\po S)\right) \quad .
\eea
Contrary to the situation in conformal gauge the matter action therefore
still contains a coupling to a zweibein component which is the 
price to pay for a simple ${\cal L}^g$. 
The generating functional of Green functions derived carefully 
from an integral in (extended) phase space for the geometric 
variables \cite{kum97} after integrating the ghosts becomes 
\begin{eqnarray}
\plabel{Zbegin}
W &= &\int ({\cal D}\sqrt{\epo}S)({\cal D}X)({\cal D}X^+)({\cal D}X^-)
({\cal D}e^+_1)({\cal D}e^-_1)({\cal D}\omega_1)F\times\nonumber \\
 && \exp\left[\frac{i}{\hbar}\int_x {\cal{L}}_{gf} \right] .
\end{eqnarray}
For the scalar field a nontrivial measure has been introduced in order
to retain invariance under general coordinate transformations 
\cite{fuj88}. We explicitely display $\hbar$ to keep track of loop 
orders in a transparent manner. 
To compute (\ref{Zbegin}) 
we {\it first} integrate over $\emo$, $X^-$ and $\omo$ to obtain 
\begin{eqnarray}
\plabel{W-half}
W &=&\int ({\cal D}\sqrt{\epo}S)({\cal D}X)({\cal D}X^+)({\cal 
D}e^+_1)\delta_{(e^-_1)}\delta_{(X^-)}\delta_{(\omo)}F\nonumber \\
&&\quad \times \exp\left[\frac{i}{\hbar}\int_x ({\cal{L}}^m_{gf} -\epo V(X) 
+{\cal L}^s)\right]
\end{eqnarray}
where the delta functions 
\begin{eqnarray}
\delta_{(\emo)}&=&\delta \left(-\pz X^++j^+ + (\pz S)^2 \right) \quad ,\\
\delta_{(X^-)}&=&\delta \left(\pz \epo +J^+ \right) \quad , \\
\delta_{(\omo)}&=&\delta \left(X^+-\pz X +j \right)\plabel{sec2e}
\end{eqnarray}
are immediately used to integrate out the remaining 
variables $X^+$,
$\epo$ and $X$.
This reduces (\ref{W-half}) to
\begin{equation}
W=\int ({\cal D}\sqrt{e^+_1}S) 
e^{\frac{i}{\hbar}\int d^2x (J^-X^++j^-\epo +JX -\epo V(X)-(\pz S)(\po S))} \quad ,
\end{equation} 
where $X^+$, $\epo$ and $X$ thus are expressed as
\begin{equation}
\begin{array}{ccccc}
X^+&=&\pzinv j^+ + \pzinv (\pz S)^2 &=&X_0^+ + \pzinv (\pz S)^2\\
\epo &=&-\pzinv J^+&{}&{} \\
X&=&\pzinv(X^+ +j)&=&X_0 + \pzinvsq (\pz S)^2 \quad .
\plabel{XeX}
\end{array}
\end{equation}
$X_0$ and $X_0^+$ represent $X$ and $X^+$ in the absence of matter fields 
(zero loop order). The (ambiguous) 
nonlocal expressions for the Green functions $\pz^{-1}$ and $\pz^{-2}$
are defined properly in the Appendix.
These integrations produce a factor $\det (\pz )^{-3}$ which cancels
exactly the Faddeev--Popov determinant (\ref{FPdet}). 
We recall that this cancellation also occurs in the presence of 
the kinetic 
dilaton term in (\ref{lfirst}) \cite{kum97}. In the present case this ghost
cancellation is not important because (\ref{FPdet}) is field independent anyhow.
To obtain the same result in conformal gauge several physical arguments
such as the vanishing of the ghost contribution to Hawking radiation 
had to be invoked by
Giddings and Strominger \cite{stro93}. It should be stressed, though,  that the 
triviality of this ghost contribution only holds in the gauge employed in our 
present paper.
Expanding $V(X)$ around $X_0$ 
\begin{eqnarray}
V(X)&=&V_{(0)}+V_{(1)}+\Delta V\; , \\
V_{(0)}&=&V(X_0)\; , \\
V_{(1)}&=&V'(X_0) \pzinvsq (\pz S)^2\; , \\
\Delta V&=&
\sum_{n=2}^{\infty}\frac{V^{[n]}(X_0)}{n!}\left(\pzinvsq (\pz S)^2 
\right)^n\; \label{deltaV} ,
\end{eqnarray} 
the matter field integration to arbitrary orders is contained in the factor
$W_S$ of 
\begin{equation}
\plabel{W-mat}
W=W_S e^{\frac{i}{\hbar}\int d^2x (J^-X_0^++j^-\epo +JX_0 -\epo V_{(0)}}) \quad , 
\end{equation}
i.e.
\begin{equation}
\plabel{sec2k}
W_S=\int ({\cal D}\sqrt{e^+_1}S) 
e^{\frac{i}{\hbar}\int d^2x  
\left( -\epo \Delta V + (\Emo (\pz S)^2 - (\pz S)(\po S)) -QS 
\right)}\; .
\end{equation}
We introduced the abbreviation 
\begin{equation}
\plabel{sec2l}
\Emo = \pzinvsq J -\pzinv J^- -\pzinvsq (\epo V'(X_0)) 
\end{equation}
in order to subsume $V_{(1)}$ into the propagator term.
$\Emo$ clearly differs from $e_1^-$  but will formally play a similar role.
Before proceeding with the remaining matter field integration we for once
look back to the matterless case.
There the generating functional for connected Green functions reduces to
\begin{equation}
\plabel{Z-no-mat}
Z=J^-\pzinv j^++J\pzinvsq j^+ +J\pzinv j -J^+\pzinv 
\left( V(\pzinvsq j^+ +\pzinv j) +j^- \right)\; .
\end{equation}
It is clear from (\ref{lfirst})  that the only possible vertex is produced by
$\epo V(X)$. It consists of one incoming $\epo$ line and multiple
$X$ lines depending on the form of the potential $V(X)$.
However, as we see from (\ref{Z-no-mat}) $X$ propagates only to 
(the sources of) $\emo$ or $\omo$.
With these propagators it is impossible to construct any loops for the 
matterless case and the absence of quantum corrections follows almost
trivially. This (for $U(X) = 0$ simple) 
perturbative argument had been  replaced by the nonperturbative path 
integral in \cite{kum97}.

Returning to (\ref{W-mat}) we still have to perform the $S$ integration.
As usual in perturbative quantum field theory the terms of higher than 
quadratic order 
in a Green function are replaced by the functional derivatives with respect 
to the sources $Q$, viz.\ $i\hbar \delta/\delta Q$.  

The integration of the term quadratic in $S$ and thus comprising the full 
propagator in the (exact!) geometric background yields 

\begin{equation}
\int ({\cal D}\sqrt{e^+_1}S)
e^{\frac{i}{\hbar} \int d^2x \Emo (\pz S)^2 - (\pz S)(\po S)-QS}=
e^{i\int d^2x S_P(\Emo ,\epo )}e^{\frac{-i}{4\hbar}\int Q\prop^{-1}Q}
\end{equation}
where $\prop^{-1}$ is defined as the inverse of the differential operator
\begin{equation}
\plabel{Gamma}
\prop = \pz \po -\pz E_1^- \pz = \partial_0\vartheta \quad .
\end{equation}
In terms of the  regularized $\partial^{-1}_{\mu}$ of the Appendix we 
define the corresponding inverse $\vartheta^{-1} = (\partial_1 - 
E_1^- \partial_0)^{-1}$.  $S_P$ denotes the Polyakov-Liouville 
action, formally written as 
\begin{equation}
\plabel{S-Pol}
S_P= \sqrt{-g} R \frac{1}{\Box} R \; ,
\end{equation}
where, however,
$R$ and $\Box$ have to be expressed in terms of $\Emo$. Thus in a 
dilaton theory with $V \neq 0$ according to (\ref{sec2l}) more types of 
external zweibein (or metric) lines are possible. From 
(\ref{sec2k}) to 
second loop order in $\Delta V$ 
\begin{eqnarray}
{}&{}& \exp {\frac{i}{\hbar}
\int (-\epo  \Delta V)}\exp {\frac{-i}{4\hbar}\int Q \prop^{-1}Q}
\nonumber \\
{}&{}& =\left( 1 - \frac{i}{\hbar}\int(\epo 
\frac{V^{''}(X_0)}{2}\left(\pz^{-2} (\pz \frac{i\hbar \delta}{\delta Q} )^2 
\right)^2 )+...\right)
e^{\frac{-i}{2\hbar}\int Q \prop^{-1}Q} |_{Q=0} \nonumber\\
{}&{}& =1+\int_x i\hbar \epo \frac{V''(X_0)}{8} \gamma (x) +O(\hbar^2)
\end{eqnarray}
follows, where we introduced the abbreviation $\gamma = \gamma 
(x), x = (x^0,x^1)$, in the last line 
which in a suggestive notation for the nonlocal expressions $\pz^{-2}$ 
is given by
\begin{eqnarray}
\label{loops}
\gamma (z) &=& \int dy dz \langle  x|\partial_0^{-2} |y\rangle \langle x|\partial_0^{-2} 
|z\rangle  \nonumber \\
{}&{}&\left(\partial_0^y \vartheta^{-1} (y,y) 
\partial_0^z\;\vartheta^{-1} (z,z) + 2 
\left(\partial_0^z\;\vartheta^{-1} (y,z)\right)^2 \right)
\end{eqnarray}

The expression $\partial_0^z\vartheta^{-1}(z,z)$ has to be understood as
$\lim_{y \to z} \partial_0^z \vartheta^{-1}(y,z)$, i.e.\ the 
differential only refers to the second argument. 
 $\langle  z |{\cal O}| z \rangle $ generically denotes  the corresponding
operator ${\cal O}$ in a $z$ representation.
Thus the two-loop contribution consists of two parts. The first one represents
two tadpole like single loops attached to the point $z$ by something resembling
a free propagator, the second term in (\ref{loops}) is a genuine two loop one
(Fig. 1).

\begin{figure}
\begin{center}
\input{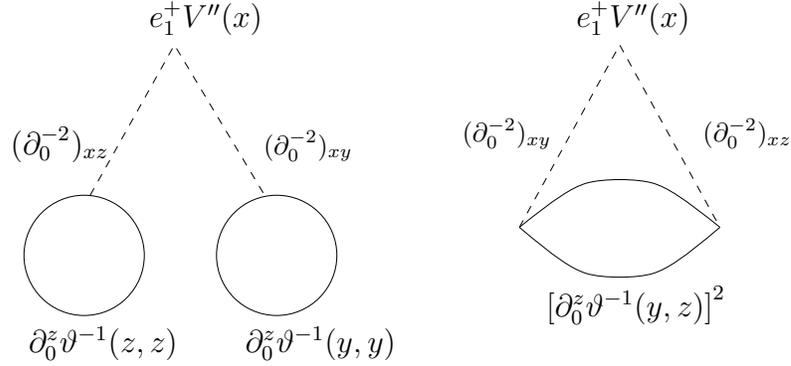}
\end{center}
\label{loopfig}
\caption{The two diagrams corresponding to (\ref{loops}).}
\end{figure}

\subsection{Two Loop Field Independence}

Before we continue the computation of the two loop results for the 
generating functional for connected Green functions we  demonstrate 
that the $\gamma$-contribution is field independent.
Both terms on the r.h.s.  of (\ref{loops}) contain the expression 
 $ \langle x|\vartheta^{-1}\partial_0 |y\rangle  $ .
Realizing that $\partial_0^{-2}$ is local in the $x_1$ coordinate
means that with our definition of $\vartheta^{-1}$ the building blocks of 
our diagrams are 

\begin{equation}
\langle x^0,x^1|\vartheta^{-1} \pz |y^0,x^1\rangle =
\langle x^0,x^1|\frac 1{\po -\Emo \pz } \pz |y^0,x^1\rangle 
\end{equation}
where $x_0$ and $x_1$ are "time" and "space" coordinates.  
The operator on the r.h.s.\ may be represented as 
\begin{eqnarray}
(P^{-1}\po  P)^{-1} & = & 
P^{-1}\po^{-1} P \nonumber \\
P(x^1) & = &{\rm {\cal P}exp } \left ( -\int^{x^1} dz^1 E_1^-(z^1)\pz \right )
\end{eqnarray}
where ${\rm {\cal P}exp}$ is the path ordered exponential satisfying the operator
identity $\po P(x^1)=P(x^1)(-E_1^-(x^1)\pz )$. 
Only $\po^{-1}$ is non-local in $x_1$. Hence we obtain 
\begin{equation}
\plabel{trick}
\langle x^0,x^1|P^{-1}(\partial_1^{-1})P\pz |y^0,x^1\rangle =
c \partial_0 \delta (x^0-y^0)
\end{equation}
where $c$ denotes the limit $x^1 \to y^1$ of the regularized 
$\langle x^1|(\partial_1^{-1})|y^1\rangle =\mp \theta (x^1-y^1)e^{\pm i\mu(x^1-y^1)}$
the sign depending on the regularization 
$\partial_1 \to \nabla_1$ or $\tilde{\nabla}_1$.
This expression, as well as $\gamma$ itself, 
 is therefore independent of sources and thus may be
absorbed in a renormalization of the potential $V$.
Actually the first term of (\ref{loops}) after taking the limit $x^0 \to y^0$
vanishes because of (\ref{trick}).

In the generating functional for connected Green functions 

\begin{eqnarray}
Z &=& \frac{\hbar}{i}\log W  \nonumber \\
\plabel{Z-loop}
{} &=& J^-X_0^++j^- \epo +JX_0 -\epo V_R(X_0)+\hbar S_P(\Emo ,\epo ) 
+O(\hbar^3)\\
V_R &=& V - \hbar^2 \gamma V'' 
\plabel{sec2r}
\end{eqnarray}
everything is expressed in terms of the sources (cf.\ (\ref{XeX}) and 
(\ref{sec2l})). 

For some time it has been argued \cite{odi92} that a generalized 
dilaton theory in d = 1+1 can be considered `renormalized' if the 
`potential' $V(X)$ by quantum corrections is changed into $V_R$ 
in (\ref{sec2r}). In those previous studies such a `renormalization' was 
supposed to occur even in the absence of matter which we have 
found to be an artefact of their approach 
by working in a more suitable gauge 
\cite{kum92}. However, in (\ref{sec2r})  this renormalization 
phenomenon --- as induced by quantum corrections from 
matter --- is a genuine one. 

Of course, only our present two-loop effect 
produces as a correction a second derivative of $V$. But let us 
assume that this remains true for higher orders (cf.\ eq.\ (\ref{deltaV})) 
leading to corresponding higher derivatives, and that $V(X)$ may be 
decomposed in a power series in $X$ 

\begin{equation}
\plabel{sec2s}
V(X)=\sum_n \frac {v_n}{n!} X^n \; .
\end{equation}

If the sum in (\ref{sec2s}) extends to infinity, each term will acquire 
quantum corrections and thus a separate normalization condition. 
Keeping in mind that the global properties of the classical 
solution depend crucially on $V$, such a renormalization 
essentially means to fix the global properties order by order, something 
not acceptable in a `renormalizable' theory of the usual 
sense.

The situation becomes even worse when  
singularities are present (poles of $V$ or noninteger powers).
This would be the case in physically more interesting models. 
Then differentiation inescapably produces 
completely new terms in each new loop order. \\
There are only two exceptions: Either a polynomial 
behaviour of $V$, where only a finite number of coefficients need to be 
redefined, or an exponential behaviour 
 $V(X)=\alpha \exp (\beta X)$ where the
renormalized potential is an exponential again
and only one parameter $\alpha$ needs to be renormalized.
Note that this potential gives black hole solutions
\cite{cru96}, sharing however with the dilaton black hole of 
\cite{man91} 
the null-completeness at the singularity \cite{kat97}.

\subsection{Effective Action}

In semiclassical calculations the effective action $\Gamma$ 
including quantum corrections \cite{man91} is used extensively. 
Starting from (\ref{Z-loop}) the `mean fields' to order $\hbar^2$  are
\begin{eqnarray}
\overline{\epo}&=\frac{\delta Z}{\delta j^-}=&\epo \; , \\
\overline{\omo}&=\frac{\delta Z}{\delta j}=&-\pzinv J+\pzinv \epo {V_R}' +
\hbar \pzinv  \left( S'_P\epo \pzinvsq  {V_R}'' \right) 
+O(\hbar^3)\; , \\
\overline{\emo}&=\frac{\delta Z}{\delta j^+}=&
\pzinvsq J-\pzinv J^--\pzinvsq 
\left(\epo {V_R}'+ \hbar S'_P\epo \pzinvsq  {V_R}'' \right) +O(\hbar^3) 
\label{bar-e}\; , \\
\overline{X}&=\frac{\delta Z}{\delta J}=&X_0+\hbar \pzinvsq S_P' +
O(\hbar^3) \label{bar-X}\; , \\
\overline{X^+}&=\frac{\delta Z}{\delta J^-}=&X_0^+ +\hbar \pzinv S_P' +
O(\hbar^3)\; .
\end{eqnarray}
$\overline{X^-}$ is not needed in the following. 
$\epo$, $X_0$, $X_0^+$ are functions of sources as in (\ref{XeX}) and
$V_R$ is a function of $X_0$ only.

Using these relations extensive cancellations occur 
 and we are left with
\begin{eqnarray}
\Gamma&=&Z-J\overline{X}-J^{\pm}\overline{X^{\mp}}-j\overline{\omega_1}
-j^{\pm}\overline{e_1^\mp}  
\nonumber \\
&=& S_{cl}(\overline{X},\overline{e},\overline{\omega}) +\hbar S_P(\Emo ,\epo)
-  \hbar^2 S_P ' \epo \pzinvsq {V_R}''(X_0) \pzinvsq S_P' +
O(\hbar^3)
\end{eqnarray}
Using (\ref{bar-e}) and (\ref{bar-X}) 
\begin{eqnarray}
\overline{\emo}=\Emo -
\hbar \pzinvsq \left( S'_P\epo \pzinvsq  {V_R}''(X_0) \right) +O(\hbar^2) \\
\overline{X}=X_0 + O(\hbar)
\end{eqnarray}
yields for the loop expansion for the generalized Polyakov term 
(\ref{S-Pol}) with (\ref{sec2l}) 
\begin{equation}
S_P(E^-_1,e^+_1)=S_P(\overline{e^-_1},\overline{e^+_1}) +
\hbar S_P ' \epo \pzinvsq {V_R}''(X_0)
\pzinvsq S_P' +
O(\hbar^2)\; .
\end{equation}
The final result is 

\begin{equation}
\Gamma=S_{cl}(\overline{e_1^\pm},\overline{\omega},\overline{X})+
\hbar S_P(\overline{e^-_1},\overline{e^+_1}) +
O(\hbar^3)
\end{equation}
where the potential $V$ is to be replaced by $V_R$ everywhere. Hence 
also in that expression the  whole 
effect of two-loop corrections is a renormalization of the dilaton
potential. It should be noted  that in the effective action to 
two-loops the corrected Polyakov term still is expressed 
 purely in terms of the mean field $\overline{\emo}$ whereas
in (\ref{Z-loop}) its two-loop corrections required the presence 
of  $\Emo$.

\section{Conclusions}
\label{sec-concl}

The crucial ingredient of our approach is the use of the
Eddington-Finkelstein gauge (or a Weyl-type gauge for the 
Cartan-variable) and a first order formulation for the action. 
In  our previous work \cite{kum97}
it helped us to calculate the effective action for pure dilaton
gravity to all orders of perturbation theory.  Here we were able 
to calculate the two--loop effective action in dilaton gravity with
matter treating the latter as a perturbation.  Thanks to a 
field-independence of the correction term in 2-loops --- which 
does not seem likely to persist in higher loop orders, though --- 
the result 
allows the interpretation as a renormalization of the potential. 
A two loop generalization of the (generalized) Polyakov term 
could be derived as well. Interestingly enough the next loop 
order in the effective action does not depend on all the 
complicated further terms found for the Polyakov corrections. 

The attentive reader will have noticed that throughout the present paper
the technique of generating functionals has been used relying heavily
on $J_i$, the sources for the $X_i$ which are nothing else but the sources
of the canonical momenta in phase space. In fact, in an expression like
(\ref{Z-loop}) the geometric part vanishes when those sources are eliminated
by naively putting $J_i=0$. As we will show in forthcoming work
a well defined procedure for eliminating $J_i$ exists yielding a generating 
functional with sources for the canonical coordinates only.
The results obtained in this way are fully consistent with those of
the second reference in \cite{kum92}, where canonical momenta were integrated
out first and no sources for them were introduced.

To 
compare with related recent results of other authors we note 
 that at one--loop order the authors of \cite{mik96} obtained
the BPP action \cite{bos95}  instead of the usual Polyakov term.
Originally the BPP action was introduced "by hand" to achieve
solvability of the effective equations of motion. 
In one-loop calculations it can appear {\it only} in the case of 
non-minimal coupling of the scalar field \cite{eli94}. A possible origin of
such a term might be that the authors \cite{mik96} assumed 
that the classical condition $\exp \rho =\exp \phi$,
connecting the dilaton and the scale factor of the metric in 
conformal gauge, can be preserved by loop corrections.\\

\appendix
\label{appA}
\section{Regularized Inverse Derivatives}

In the preceding sections we frequently encountered inverse derivative operators.
Here we shall define a proper infrared regularization scheme and list
the corresponding calculation rules which where used in the main text.
We introduce two regularized Green functions $\nabla_0^{-1}$ and 
$\tilde{\nabla}_0^{-1}$ to replace $\partial_0^{-1}$. 
\begin{equation}
\pz^{-1} \Rightarrow
\begin{cases}
 \lim_{\mu \to 0}\left(\pz -i\mu \right)^{-1}=
 \lim_{\mu \to 0}\left(\nabla_0^{-1} \right) \\
 \lim_{\mu \to 0}\left(\pz +i\mu \right)^{-1}=
 \lim_{\mu \to 0}\left(\tilde{\nabla}_0^{-1} \right) \quad
\end{cases}
\end{equation}
where $\mu =\mu_0 -i\varepsilon$. $\mu_0 \to +0$ represents the IR regularization, 
proper asymptotic behavior (cf. (\ref{inv}),(\ref{inv2}) below) is 
provided by $\varepsilon \to +0$.
 Note that a partial integration transforms
 $\nabla_0^{-1}$ into  $\tilde{\nabla}_0^{-1}$ and also that 
$\tilde{\nabla}_0^{-1}$ is not the complex conjugate of $\nabla_0^{-1}$.
The inverse operators are defined as the Green functions  $\nabla_0$ and
$\tilde{\nabla}_0$ and are calculated straightforwardly
\begin{eqnarray}
\label{inv}
\left( \nabla_0^{-1} \right)_{x,y}&=& -\theta (y-x)e^{i\mu (x-y)} \\
\label{inv2}
\left(\tilde{\nabla}_0^{-1}\right)_{x,y}&=& \theta (x-y)e^{-i\mu (x-y)} \quad ,
\end{eqnarray}
where $\theta$ denotes the step function.
The inverse squared operators are defined as the Green functions of
$(\nabla_0)^2$ and $(\tilde{\nabla}_0)^2$
and are given by
\begin{eqnarray}
\left( \nabla_0^{-2} \right)_{x,y}&=& (y-x)\theta (y-x)e^{i\mu (x-y)} \\
\label{invsq}
\left(\tilde{\nabla}_0^{-2}\right)_{x,y}&=& (x-y)\theta (x-y)e^{-i\mu (x-y)}\quad .
\end{eqnarray}
Using (\ref{inv}) to (\ref{invsq}) the following rules may be verified 
easily 
\begin{alignat}{2}
%\begin{tabular}{ll \TVR42}
&\nabla_0 \nabla_0^{-2} =\nabla_0^{-1} & \qquad &
\tilde{\nabla}_0 \tilde{\nabla}_0^{-2} =\tilde{\nabla}_0^{-1}  \\
&\nabla_0 \tilde{\nabla}_0^{-2}= \tilde{\nabla}_0^{-1} -2i\mu \tilde{\nabla}_0^{-2} & \qquad  &
\tilde{\nabla}_0 \nabla_0^{-2}= \nabla_0^{-1} +2i\mu \nabla_0^{-2}  \\
&\nabla_0 \tilde{\nabla}_0^{-1}=\delta(x-y) -2i\mu \tilde{\nabla}_0^{-1}  & \qquad &
\tilde{\nabla}_0 \nabla_0^{-1}=\delta(x-y) +2i\mu \nabla_0^{-1} \quad .
\end{alignat}

\noindent
Note that in the main text only these types of operations appeared 
and therefore
the limit $\mu \to 0$ does not cause any divergencies, because 
$\mu$ does not appear with negative powers. Therefore this  
regularization scheme seems much superior to the one used in 
\cite{kum92}. 

\section*{Acknowledgement}

This work has been supported by Fonds zur F\"orderung der
wissenschaftlichen For\-schung (FWF) Project No.\ P 10221--PHY.  One
of the authors (D.V.) thanks GRACENAS and
the Russian Foundation for 
Fundamental Research, grant
97-01-01186, for financial support.

\vfill

\end{document}